
\input amstex
\documentstyle{amsppt}
\NoRunningHeads
\NoBlackBoxes
\hyphenation{Ko-lo-kol-tsov}
\magnification=1200
\define\slt{\operatorname{sl}(2,\Bbb C)}
\define\SU{\operatorname{SU}}
\define\Vrt{\operatorname{Vert}}

\define\sgn{\operatorname{sgn}}
\define\V{W}

\define\tr{\operatorname{tr}}
\define\ttt{\operatorname{T}}
\define\ti{\operatorname{int}}
\define\e{\operatorname{ext}}

\topmatter
\title
Belavkin--Kolokoltsov watch--dog effects in interactively controlled
stochastic computer-graphic dynamical systems. A summary of mathematical
researches.
\endtitle
\author Denis V. Juriev
\endauthor
\address Mathematical Division, Research Institute for System Studies,
Russian Academy of Sciences, Moscow, Russia
\endaddress
\email juriev\@systud.msk.su
\endemail
\address
Laboratoire de Physique Th\'eorique de l'\'Ecole Normale Sup\'erieure,
24 rue Lho\-mond, 75231 Paris Cedex 05, France (1.XII.1993--1.XII.1994)
\endaddress
\email juriev\@physique.ens.fr
\endemail
\abstract This paper contains a summary of mathematical researches
of stochastic properties of the long time behavior of a continuously
observed (and interactively controlled) quantum--field top.
Applications to interactively controlled stochastic computer-graphic
dynamical systems are also discussed.
\endabstract
\endtopmatter
\document
\head I. INTRODUCTION\endhead

The main difficulty to account the high--frequency eye tremor in {\it
mobilevision\/} ({\it MV\/}) [Ju2, Ju3, Ju4, Ju5] is that in this case a
solution of the complete MV evolution equations in real time requests about
10$^8$--10$^9$ arithmetical operations per second (moreover, it needs special
displays of a high refreshing rate ($\sim$ 300--500 frames per second) and a
small image inertia). Such account may be performed only on a narrow class of
computers for the purposes of scientific experiments on peculiarities of human
vision in interactive computer-graphic systems [Ju2, Ju3], but it is very
inconvenient for an assimilation of MV as a computer-graphic tool f.e. for an
interactive visualization of 2D quantum field theory [Ju5]. So one should to
use some stochastic simulation of the interactive processes, i.e. to consider
an imitated stochastic process instead of the tremor. Such approach leads to
{\it stochastic mobilevision\/} ({\it SMV\/}) [Ju2], which evolution equations
have a stochastic Belavkin--type form [Be, BK, Ko1]. It seems that the
interactive effects for ordinary MV and SMV are similar in general, because
the interactive processes accounting saccads are not stochastized; though it
is not an undisputable fact that they are always identical (f.e. in a
situation of the so--called "lateral vision"). The combination of MV with
cluster and spline techniques allows to work on computers with 10$^5$--10$^6$
arithmetical operations per second (as well as to use simpler devices for eye
motion detection and a wider class of displays), whereas all enumerated above
circumstances make the tremor accounting in terms of ordinary MV almost
unreasonable nowadays.

Nevertheless, all these advantages of SMV are not crucial in view of the
permanent progress in the computer hightech (for example, the using of a
distributed parallel processing allows to diminish the request for the tremor
accounting ordinary MV to $\sim$ 10$^6$ arithmetical operations per second,
etc.). A deeper advantage of SMV is more fundamental --- it is a presence of
the Belavkin--Kolokoltsov watch--dog effects ([Ko2], see also the original
papers [CSM, MS] where "watch--dog effects" or a "quantum Zeno paradox" were
put under a consideration) in SMV in certain rather natural and general cases
(i.e. for certain values of internal parameters measuring the degree of
localization of interaction) that means an {\it a priori\/} finiteness of
sizes of stochastic "cores" of an image during observation, moreover they may
be diminished to several pixels by a suitable choice of a {\it free\/}
controlling parameter (the so--called {\it "accuracy of measurement"\/} [Be,
BK, Ko1, Ko2]). The watch--dog effect may be considered as a weaker but also
tamer form of nondemolition than the quasistationarity [Ju2, Ju3]: there
exists a wide class of models, in which the first is observed whereas the
least is broken, one may consider canonical projective $G$--hypermultiplets
[Ju3] (see also [Ju4]) as a simple example.

Thus, a transition from MV to SMV partially solves {\it a priori\/} the main
problem of dynamics in interactive psychoinformation computer-graphic systems
[Ju2, Ju3] --- a problem of the nondemolition of images by the interactive
processes (i.e. their stability under observation). Certainly, SMV does not
solve the nondemolition problem completely {\it a priori}. It only garantees
that the stochastic cores of image have finite sizes during observation, it
means that details of image do not diffuse. Nevertheless, they may move, being
ruled by the slow eye movements. So though details of image are perserved, the
image may be destructed as a whole. It seems that the quasistationarity
conditions [Ju2, Ju3] are realistic complements to watch--dog effects and
together they provide a complete long--time nondemolition of images.

Also it should be marked that such {\it a priori\/} nondemolition in SMV
confirms a presence of {\it a posteriori\/} one in the tremor accounting
ordinary MV.

The purpose of this note is to investigate the Belavkin--Kolokoltsov
watch--dog effects in SMV mathematically.

Summarizing arguments above one may conclude that such investigations are
motivated by the overlapping of two problems:
\roster
\item"1)" the difficulty to account the high--frequency eye tremor in ordinary
mobilevision, which leads to the necessity to consider tremor's stochastic
simulations;
\item"2)" the main problem of dynamics in interactive psychoinformation
computer-graphic systems, i.e. a problem of the nondemolition of images by the
interactive processes; it motivates investigations of long--time properties of
(nonlinear) stochastic dynamics in SMV.
\endroster
So the first problem explains, why stochastic mobilevision is put under a
consideration, the second one explains a choice of questions, which are
tried to be solved in the paper.

\head II. MATHEMATICAL ASPECTS OF MOBILEVISION [Ju4, Ju5]\endhead

This paragraph is devoted to a brief exposition of mathematical (geometric)
aspects of ordinary mobilevision. It may be omitted by an educated reader.

\eightpoint \subhead 2.1. Interpretational geometry and anomalous virtual
realities \endsubhead Interpretational geometry is a certain geometry related
to interactive computer-graphic psychoinformation systems. Mathematical data
in such systems exist in the form of an interrelation of an interior geometric
image (figure) in the subjective space of observer and an exterior
computer-graphic representation; the least includes the visible elements
(draws of figure) as well as of the invisible ones (f.e. analytic expressions
and algorythms of the constructing of such draws). Process of the
corresponding of a geometrical image (figure) in the interior space of
observer to a computer-graphic representation (visible and invisible elements)
is called {\it translation}; the visible object maybe nonidentical to the
figure, in this case partial visible elemnts may be regarded as modules, which
translation is realized separately; the translation is called by {\it
interpretation\/} if the translation of partial modules is realized depending
on the result of the translation of preceeding ones.

\definition{Definition 1} A figure, which is obtained as a result of the
interpretation, is called {\it interpretational figure\/}; the draw of an
interpretational figure is called {\it symbolic\/}.\enddefinition

Note that the simbolic draws may be characterized only as "visual
perception technology" of figure but not as its "image".

The computer--geometric description of mathematical data in interactive
information systems is deeply related to the concept of anomalous virtual
reality. It should be mentioned that there exist several approaches to
foundations of geometry: in one of them the basic geometric concept is a space
(a medium, a field), geometry describes various properties of a space and its
states, which are called the draws of figures; it is convenient to follow this
approach for the purposes of the describing of geometry of interactive
information systems; the role of the medium is played by an anomalous virtual
reality, the draws of figures are its certain states.

\definition{Definition 2}

{\bf A.} {\it Anomalous virtual reality\/} ({\it AVR\/}) {\it in a narrow
sense\/}
is a certain system of rules of non--standard descriptive geometry adopted to
a realization on videocomputer (or multisensor system of "virtual reality"
[BC, Rh, VR1, VR2, VR3]); {\it anomalous virtual reality in a wide sense\/}
contains also an image in the cyberspace made accordingly to such system of
rules; we shall use this term in a narrow sense below.

{\bf B.} {\it Naturalization\/} is the corresponding of an AVR to an abstract
geometry or a physical model; we shall say that the AVR {\it naturalizes\/}
the model and such model {\it transcendizes\/} the naturalizing AVR. {\it
Visualization in a narrow sense\/} is the corresponding of certain images or
visual dynamics in the AVR to objects of the abstract geometry or processes in
the physical model; {\it visualization in a wide sense\/} also includes the
preceeding naturalization.

{\bf C.} An anomalous virtual reality, whose images depends on an observer, is
called {\it intentional anomalous virtual reality\/} ({\it IAVR\/});
generalized perspective laws in IAVR contain the equations of dynamics of
observed images besides standard (geometric) perspective laws; a process of
observation in IAVR contains a physical process of observation and a virtual
process of intention, which directs an evolution of images accordingly to
dynamical laws of perspective. \enddefinition

In the intentional anomalous virtual reality objects of observation present
themselves being connected with observer, who acting on them in some way,
determines, fixes their observed states, so an object is thought as a
potentiality of a state from the defined spectrum, but its realization depends
also on observer; the symbolic draws of interpretational figures are presented
by states of a certain IAVR.

Note that a difference of descriptive geometry of computer-graphic
information systems from the classical one is the presense of colors as
important bearers of visual information; a reduction to shape graphics, which
is adopted in standard descriptive geometry, is very inconvenient, since the
use of colors is very familiar in the scientific visualization [SV1, SV2, Vi1,
Vi2]. The approach to the computer-graphic interactive information systems
based on the concept of anomalous virtual reality allows to consider an
investigation of structure of a color space as a rather pithy problem of
descriptive geometry, because such space maybe much larger than the usual one
and its structure may be rather complicated.

\definition{Definition 2D}
A set of continuously distributed visual characteristics of image in an
anomalous virtual reality is called {\it anomalous color space\/}; elements
of an anomalous color space, which have non--color nature, are called {\it
overcolors\/}, and quantities, which transcendize them in the abstract model,
are called {\it "latent lights"\/}. {\it Color--perspective system\/} is a
fixed set of generalized perspective laws in fixed anomalous color space.
\enddefinition

\subhead 2.2. Quantum projective field theory and mobilevision\endsubhead It
seems to be a significant fact that 2D quantum field theory maybe expressed in
terms of interpretational geometry, so that various objects of this theory are
represented by interpretational figures. The keypoint is {\it
mobilevision\/} ({\it MV\/}), which is an IAVR naturalizing {\it the quantum
projective field theory\/} ({\it QPFT\/}; [BR, Ju3] and refs wherein); the
process of naturalization is described in [Ju2, Ju3, Ju4, Ju5].

Let's concentrate our attention on the basic concepts of the QPFT, which
naturalization mobilevision is.

\definition{Definition 3A} {\it QFT--operator algebra\/} ({\it
operator algebra of the quantum field theory, vertex operator algebra, vertex
algebra\/}) is the pair $(H,t^k_{ij}(\vec x))$: $H$ is a linear space,
$t^k_{ij}(\vec x)$ is $H$--valued tensor field such that $t^l_{im}(\vec
x)t^m_{jk}(\vec y)=t^m_{ij}(\vec x-\vec y)t^l_{mk}(\vec y)$.
\enddefinition

Let us intruduce the operators $l_{\vec x}(e_i)e_j=t^k_{ij}(\vec x)e_k$, then
the following relations will hold: $l_{\vec x}(e_i)l_{\vec
y}(e_j)=t^k_{ij}(\vec x-\vec y)l_{\vec y}(e_k)$ ({\it operator product
expansion\/}) and $l_{\vec x}(e_i)l_{\vec y}(e_j)=l_{\vec y}(l_{\vec x-\vec
y}(e_i)e_j)$ ({\it duality\/}).
Also an arbitrary QFT--operator algebra one can define an operation depending
on the parameter: $m_{\vec x}(e_i,e_j)=t^k_{ij}(\vec x)e_k$; for this
operation the following identity holds: $m_{\vec x}(\cdot,m_{\vec
y}(\cdot,\cdot))=m_{\vec y}(m_{\vec x-\vec y}(\cdot,\cdot),\cdot)$; the
operators $l_{\vec x}(f)$ are the operators of the left multiplication in the
obtained algebra.

\definition{Definition 3B}
QFT--operator algebra $(H,t^k_{ij}(u); u\in\Bbb C)$ is called {\it (derived)
QPFT--operator
algebra\/}
iff (1) $H$ is the sum of Verma modules $V_{\alpha}$ over $\slt$ with the
highest vectors $v_{\alpha}$ and the highest weights $h_{\alpha}$, (2)
$l_u(v_{\alpha})$ is a primary field of spin $h_{\alpha}$, i.e.
$[L_k,l_u(v_\alpha)]=(-u)^k(u\partial_u+(k+1)h_{\alpha})l_u(v_{\alpha})$,
where $L_k$ are the $\slt$ generators ($[L_i,L_j]=(i-j)L_{i+j}$,
$i,j=-1,0,1$),
(3) the (derived) rule of descendants generation holds
($[L_{-1}l_u(f)]=l_u(L_{-1}f)$).
(Derived) QPFT--operator algebra $(H,t^k_{ij}(u))$ is
called {\it projective $G$--hypermultiplet\/}, iff the group $G$ acts in it by
automorphisms, otherwords, the space $H$ possesses a structure of the
representation of the group $G$, the representation operators commute with the
action of $\slt$ and $l_u(T(g)f)=T(g)l_u(f)T(g^{-1})$.
\enddefinition

The linear spaces of the highest vectors of the fixed weight form
subrepresentations of $G$, which are called {\it multiplets\/} of projective
$G$-hypermultiplet.

Now let's describe the key moments of the process of naturalization of the
QPFT which is resulted in MV. Unless the abstract model (QPFT) has a quantum
character the images in its naturalization (MV) are classical; the transition
from the quantum field model to classical one is done by standard rules,
namely, the classical field with Taylor coefficients $|a_k|^2$ is corresponded
to the element $\sum a_k L_{-1}^k v_{\alpha}$ of the QPFT--operator algebra.
Under the naturalization three classical fields are identified with fields of
three basic colors (red, green and blue), other fields with fields of
overcolors; there are pictured only the color characteristics for the fixed
moment of time on the screen of the videocomputer as well as the perception of
the overcolors by an observer is determined by the intentional character of
the AVR of mobilevision. Namely, during the process of the evolution of the
image, produced by the observation, the vacillations of the color fields take
place in accordance to the dynamical perspective laws of MV (Euler formulas or
Euler--Arnold equations). These vacillations depend on the character of an
observation (f.e. an eye movement or another dynamical parameters); the
vacillating image depends on the distribution of the overcolors, that allows
to interpret the overcolors as certain {\it interactive vacillations\/} of the
ordinary colors. So the overcolors of MV are vacillations of the fixed type
and structure of ordinary colors with the defined dependence on the parameters
of the observation process; the transcending "latent lights" are the quantized
fields of the basic model of the QPFT.

The presence of the $\SU(3)$--symmetry of classical color space allows to
suppose that the QPFT--operator algebra of the initial model is the projective
$\SU(3)$--hypermultiplet.

\subhead 2.3. Quantum conformal and $q_R$--conformal field theories;
quantum--field analogs of Euler--Arnold tops\endsubhead
\definition{Definition 4A}
The highest vector $T$ of the weight 2 in the QPFT--opeartor algebra will
be called {\it the conformal stress--energy tensor\/} if
$T(u):=l_u(T)=\sum L_k
(-u)^{k-2}$, where the operators $L_k$ form the Virasoro algebra:
$[L_i,L_j]=(i-j)L_{i+j}+\frac{i^3-i}{12} c\cdot I$.
The set of the highest vectors $J^\alpha$ of the weight 1 in the
QPFT--operator algebra will be called the set of {\it the affine currents\/}
if $J^\alpha(u):=l_u(J^\alpha)=\sum J^\alpha_k(-u)^{k-1}$, where the
operators $J^\alpha_k$ form {\it the affine Lie algebra\/}:
$[J^\alpha_i,J^\beta_j]=c^{\alpha\beta}_\gamma
J^\gamma_{i+j}+k^{\alpha\beta}\cdot i\delta(i+j)\cdot I$.
\enddefinition

If there is defined a set of the affine currents in the QPFT--operator algebra
then one can construct the conformal stress--energy tensor by use of Sugawara
construction. Below we shall be interested in the special deformations of the
quantum conformal field theories in class of the quantum projective ones,
which will be called {\it quantum $q_R$--conformal field theories\/}; the
crucial role is played by so--called {\it Lobachevskii algebra\/} in their
constructions. In the Poincare realization of the Lobachevskii plane (the
realization in the unit disk) the Lobachevskii metric maybe written as
$w=q_R^{-1}\,dzd\bar{z}/(1-|z|^2)^2$; one can construct the $C^*$--algebra
(Lobachevskii algebra), which maybe considered as a quantization of such
metric, namely, let us consider two variables $t$ and $t^*$, which obey the
following commutation relations: $[tt^*,t^*t]=0$,
$[t,t^*]=q_R(1-tt^*)(1-t^*t)$ (or in an equivalent form $[ss^*,s^*s]=0$,
$[s,s^*]=(1-q_Rss^*)(1-q_Rs^*s)$, where $s=(q_R)^{-1/2}t$); one may realize
such variables by bounded operators in the Verma module over $\slt$ of the
weight $h=\frac{q_R^{-1}+1}2$ (this relation between $h$ and $q_R$ will be
presupposed below); if such Verma module is realized in polynomials of one
complex variable $z$ and the action of $\slt$ has the form $L_{-1}=z$,
$L_0=z\partial_z+h$, $L_1=z(\partial_z)^2+2h\partial_z$, then the variables
$t$ and $t^*$ are represented by tensor operators $D=\partial_z$ and
$F=z/(z\partial_z+2h)$. These operators are bounded if $q_R>0$ and therefore
one can construct a Banach algebra generated by them and obeying the
prescribed commutation relations; the structure of $C^*$--algebra is rather
obvious: an involution $*$ is defined on generators in a natural way, because
the corresponding tensor operators are conjugate to each other.

\definition{Definition 4B}
The highest vector $T$ of the weight 2 in the QPFT--operator algebra will
be called {\it the $q_R$--conformal stress--energy tensor\/} if
$T(u):=l_u(T)=\sum L_k (-u)^{k-2}$, where the operators $L_k$ form the
$q_R$--Virasoro algebra: $[L_i,L_j]=(i-j)L_{i+j}$ ($i,j\ge -1$; $i,j\le 1$),
$[L_2,L_{-2}]=H(L_0+1)-H(L_0-1)$,
$H(t)=t(t+1)(t+3h-1)^2/((t+2h)(t+2h-1))$ (cf.[Ro]).
The set of the highest vectors $J^\alpha$ of the weight 1 in the
QPFT--operator algebra will be called the set of {\it the $q_R$--affine
currents\/} if $J^\alpha(u):=l_u(J^\alpha)=\sum J^\alpha_k(-u)^{k-1}$, where
the operators $J^\alpha_k$ form {\it the $q_R$--affine Lie algebra\/}:
$J^{\alpha}_k=J^{\alpha}T^{-k}f_k(t)$,
$[J^{\alpha},J^{\beta}]=c^{\alpha\beta}_{\gamma}J^{\gamma}$,
$Tf(t)=f(t+1)T$,
$[T,J^{\alpha}]=[f(t),J^{\alpha}]=0$,
$f_k(t)=t\ldots(t-k),\text{ if } k\ge 0,\text{ and }
((t+2h)\ldots (t+2h-k+1))^{-1},\text{ if } k\le 0$.
\enddefinition

It should be mentioned that $q_R$--affine currents and $q_R$--conformal
stress--energy tensor are just the $\slt$--primary fields in the Verma module
$V_h$ over $\slt$ of spin 1 and 2, respectively;
if such module is realized as before then
$J_k=\partial_z^k$, $J_{-k}=z^k/(\xi+2h)\ldots(\xi+2h+k-1)$;
$L_2=(\xi+3h)\partial^2_z$, $L_1=(\xi+2h)\partial_z$, $L_0=\xi+h$,
$L_{-1}=z$,
$L_{-2}=z^2\frac{\xi+3h}{(\xi+2h)(\xi+2h+1)}$, $\xi=z\partial_z$.
So the generators $J_k^\alpha$ of $q_R$--affine algebra maybe represented via
generators of Lobachevskii
$C^*$--algebra:
$J^\alpha_k=J^\alpha t^k,\text{ if } k\ge 0,\text{ and }
J^\alpha(t^*)^{-k},\text{ if } k\le 0$,
($[J^\alpha,J^\beta]=c^{\alpha,\beta}_{\gamma}J^{\gamma}$).
That means that $q_R$--affine algebra admits a homomorphism in a tensor
product of the universal envelopping algebra $\Cal U(\frak g)$ of the Lie
algebra $\frak g$, generated by $J^\alpha$, and Lobachevskii algebra.
The (derived) QPFT--operator algebras generated by $q_R$--affine currents are
called canonical projective $G$--hypermultiplets.
The primary fields $V_k(u)=\exp(k(Q+R(\int V_1(u)\,du)))$
($R(u^n)=-\sgn(n)u^n$, i.e. $R$ is the Hilbert transform
$f(\exp(it))\mapsto-\frac{i}{2\pi}\int f(\exp(i(t-s)))\cot(s/2)\,ds$; a charge
$Q$ is defined as $Q(z^n)=\sum_{j=0}^{n-1}(j+2h)^{-1}z^n$; [BJ1,By])
of non--negative integer
spins $k$ in the Verma module $V_h$, which form a closed QPFT--operator
algebra (a subalgebra of $\Vrt(\slt)$ [BJ2], generated by vertex operator
fields $B_k(u;\nabla_h)$ [Ju1]), are not mutually local.
It is interesting to calculate $T$--exponent and monodromy of $q_R$--affine
current; it maybe easily performed by a perturbation of simple formulas for
such objects for a singular part of a current, as it was stated in [Ju2] such
perturbation by a regular part does not change the resulting monodromy.

Let $H$ be an arbitrary direct sum of Verma modules over $\slt$ and $P$ be a
trivial fiber bundle over $\Bbb C$ with fibers isomorphic to $H$; it should be
mentioned that $P$ is naturally trivialized and possesses a structure of
$\slt$--homogeneous bundle. A $\slt$--invariant Finsler connection
$A(u,\dot u)$ in $P$ is called an angular field; angular field
$A(u,\dot u)$ may be expanded by $(\dot u)^k$, the coefficients of
such expansion are just $\slt$--primary fields; the equation
$\dot\Phi_t=A(u,\dot u)\Phi_t$, where $\Phi_t$ belongs to $H$ and
$u=u(t)$ is the function of scanning, is a quantum--field analog of the Euler
formulas; such analog describes an evolution of MV image under the observation
(more rigorously, such evolution is defined in the dual space $H^*$ by
formulas $\dot\Phi_t=A^{\ttt}(u,\dot u)\Phi_t$). Regarding
canonical projective $G$--hypermultiplet we may construct a quantum field
analog of the Euler--Arnold equation $\dot A=\{\Cal H,A_t\}$, where an
angular field $A(u,\dot u)$ is considered as an element of the canonical
projective $G$--hypermultiplet being expanded by $\slt$--primary fields of
this hypermultiplet, $\Cal H$ is the quadratic $\SU(3)$--invariant element of
$S(\frak g)$, $\{\cdot,\cdot\}$ are canonical Poisson brackets in $S(\frak
g)$.  It is possible to combine Euler--Arnold equations with Euler formulas to
receive the complete dynamical perspective laws of the MV.  The main feature
of these laws is their {\it projective invariance}, so they define a natural
generalization of ordinary descriptive geometry for an interpretational case.
The projective invariance fixes Euler formulas uniquely, whereas it allows, of
course, to change Euler--Arnold equations to other ones (hamiltonian or ever
nonhamiltonian). However, such equations should provide $\SU(3)$--invariance
of the dynamical perspective laws.

\subhead 2.4. Organizing MV cyberspace \endsubhead MV cyberspace consists of a
space of images $V_I$ with the fundamental length (a step of the lattice)
$\Delta_I$ and a space of observation $V_O$ with the fundamental length
$\Delta_O$; the space of images $V_I$ is one where pictures are formed,
whereas the space of observation $V_O$ is used for a detection of eye motions;
it is natural to claim that $\Delta_O\ll A_{\tr}$, where $A_{\tr}$ is an
amplitude of the eye tremor, as well as $\Delta_I\gg\Delta_O$. The Euler
formulas maybe written as $\dot\Phi_t=A(u,\dot u)\Phi_t$, $A(u,\dot u)$ maybe
considered approximately in the form $M_1(t)\dot uV_1(u)+M_2(t)\dot u^2
V_2(u)+M_3(t)\dot u^3 V_3(u)$, where $\Phi_t\in H$ (or in the form
$\dot\Phi_t=A^{\ttt}(u,\dot u)\Phi_t$, $\Phi_t\in H^*$); here $M_i(t)$ are
data of Euler--Arnold top, $V_i(u)$ are $\slt$--primary fields in the Verma
module $V_h$, which maybe written as
$V_i(u)=(-u)^{-i}(\V_i(u)+\V^*_i(u)-D^i_0)=\sum_{j\in\Bbb Z}
(-u)^{-i-j}D^i_j$, $\V_i(u)=\sum_{j\ge 0} (-u)^{-j}D^i_j$,
$\V^*_i(u)=\sum_{j\ge 0} (-u)^jD^i_{-j}$, $D^i_{-j}=(D^i_j)^*$, where $*$ is
the conjugation in the unitarizable Verma module. The tensor operators $D^i_k$
($k\ge 0$, $i=1,2,3$) have the form $D^i_k=P_{i,k}(z\partial_z)\partial^k_z$,
$P_{1,k}(t)=1$, $P_{2,k}(t)=t+(k+1)h$,
$P_{3,k}(t)=t^2+((k+2)h+k/2)t+h(2h+1)(k+1)(k+2)/6$. It should be mentioned
that the fields $\V^{\ttt}_i(u)$ in local $\slt$--modules $V^*_h$ are defined
by rather simple expressions: $$ \align
\V^{\ttt}_1(u)=&\frac1{1-u^{-1}x},\qquad
\V^{\ttt}_2(u)=-\frac{x}{1-u^{-1}x}\partial_x+\frac{h}{(1-u^{-1}x)^2},\\
\V^{\ttt}_3(u)=
&\frac{x^2}{1-u^{-1}x}\partial^2_x-(2h+1)\frac{x}{(1-u^{-1}x)^2}\partial_x+
\frac{h(2h+1)}{3}\frac{1}{(1-u^{-1}x)^3}. \endalign $$ The matrix
$A^{\ttt}(u,\dot u)$ of size $(N,N)$ ($N$ is a number of points of $V_I$)
should be expanded in a sum of three terms $M_i(t)\dot u^iV_i(u)$ ($i=1,2,3$),
where $V^{\ttt}_i(u)$ are matrices of size $(N,N)$, depending on parameter
$u$; this parameter may have $M$ different values ($M$ is number of points of
$V_O$). Matrices $\V^{\ttt}_i(u)$ are easily calculated, one should obtain the
complete matrices $V^{\ttt}_i(u)$ making a conjugation in the unitary local
$\slt$--module $V^*_h$. Derivatives should be replaced by differences
everywhere in a standard way. Formulas for $M_i(t)$ maybe received from [MF].

\subhead 2.5. Non--Alexandrian geometry of mobilevision \endsubhead It should
be marked that almost all classical geometries use a certain postulate, which
we shall call Alexandrian, but do not include it in their axiomatics
explicitely. A precise formulation of this postulate is given below.

\proclaim{Alexandrian postulate} Any statement holding for a certain geometric
configuration remains true if this configuration is considered as a
subconfiguration of any its extension.
\endproclaim

Alexandrian postulate means that an addition of any subsidiary objects to a
given geometric configuration does not influence on this configuration.
It is convenient to describe a well--known example of non--Alexandrian
geometry (which maybe called Einstein geometry).

\remark{Example of non-Alexandrian geometry} Objects of geometry are weighted
points and lines.
Weigh-\linebreak ted points are pairs (a standard point on a plane, a real
number). They
define a (singular) metric on a plane via Einstein--type equations $R(x)=\sum
m_\alpha\delta(x-x_\alpha)$, where $(x_\alpha,m_\alpha)$ are weighted points
and $R(x)$ is a scalar curvature.
Lines are geodesics for this metric.
The basic relation is a relation of an incidence.
\endremark

It can be easily shown that Alexandrian postulate doesn't hold for such
geometry, which contains a standard Euclidean one (extracted by the condition
that all "masses" $m_\alpha$ are equal to 0).

Kinematics and process of scattering of figures maybe illustrated by another
important example of non--Alexandrian geometry --- geometry of solitons
[ZMNP]. The basic objects of KdV--soliton geometry are moving points on a
line; a configuration of such points defines a $n$--soliton solution of
KdV--equation $u_t=6uu_{xx}-u_{xxx}$ by the formulas
$u(x,t)=-2(\log(\det(E+C)))_{xx}$, where $C_{nm}=
c_n(t)c_m(t)\exp(-(\varkappa_n+\varkappa_m)t)/(\varkappa_n+\varkappa_m)$,
$c_n(t)=c_n(0)\exp(4\varkappa^3_nt)$; such solution is asymptotically free,
i.e. maybe represented as a sum of 1--soliton solutions (solitons) whereas
$t\to\pm\infty$. Soliton has the form
$u(x,t)=-2\varkappa^2\cosh^{-2}(\varkappa(x-4\varkappa^2t-\varphi))$, where
phase $\varphi$ is an initial position of soliton and $v=4\varkappa^2$ is its
velocity; scattering of solitons is two--particle, the shift of phases is
equal to
$\varkappa^{-1}_1\log|(\varkappa_1+\varkappa_2)|/|(\varkappa_1-\varkappa_2)|$
for the first (quick) soliton and
$-\varkappa_2^{-1}\log|(\varkappa_1+\varkappa_2)|/|(\varkappa_1-\varkappa_2)|$
for the second (slow) one. All examples of soliton geometries confirm the
opinion that a breaking of the Alexandrian postulate is generated by an
interaction of geometrical objects, in particular, such interaction maybe
defined by a nonlinear character of their evolution.

Let's consider now an interpretational scattering. As it was stated below a
figure in interpretational geometry is described by a pair
$(\Phi^{\ti},\Phi^{\e})$, where $\Phi^{\ti}$ is an interior image in the
subjective space of observer and $\Phi^{\e}$ is its exterior computer-graphic
draw; $\Phi^{\ti}$ is a result of interpretation of $\Phi^{\e}$. It is natural
to suppose that $\Phi^{\ti}$ depends on $\Phi^{\e}$ functionally
$\Phi^{\ti}_t=\Phi^{\ti}\left[\Phi^{\e}_{\tau\le t}\right]$ and as a rule
nonlinearly; moreover, if $\Phi^{\e}$ is asymptotically free then $\Phi^{\ti}$
is also asymptotically free. Thus, a nontrivial scattering of interacting
interpretational figures exists (i.e. although we do not know an explicit form
of dynamical equations for $\Phi^{\ti}$, their solutions, nevertheless, in
view of our assumptions maybe considered as {\it a priori\/} soliton--like),
so interpretational geometries maybe considered as non--Alexandrian ones; it
should be specially marked that the breaking of Alexandrian postulate is
realized on the level of figures themselves, but it is not observed on one of
their draws.  \tenpoint

Informatic aspects of mobilevision are considered in the second part of [Ju5].

\head III. MATHEMATICAL ASPECTS OF STOCHASTIC MOBILEVISION [Ju8]\endhead

\subhead 3.1. Mathematical set up\endsubhead
First of all, stochastic mobilevison as well as ordinary mobilevision are
interactive computer-graphic systems, the evolution of images in which is
governed by the eye movements in accordance to the certain {\it dynamical
perspective laws}, i.e. dynamical equations, which govern an evolution of
image during observation (see par.II or [Ju4,Ju5]). So their definitions are
just the specifications of such laws (it should be specially stressed that we
restrict now our interest in interactive computer-graphic systems by an
intrinsic {\it constructive\/} point of view [KT], considering them {\it as
such\/} but not as {\it descriptive\/} tools of any use for modelling or
visualizing of various physical processes (as in [Ju5]), such approach may be
rather narrow but effective and it is reasonable to adopt it for the further
discussion). The laws for MV were written in par.II (or in [Ju2, Ju3, Ju4,
Ju5]). Stochastic mobilevision have the slightly different laws. A difference
may be briefly summarized in the following terms: (1) the high--frequency eye
tremor is decoupled from the slow eye motions (including saccads), (2) it is
stochastized in such a way that it may be considered as {\it purely internal
process\/} in the system so that (3) its characteristics are not completely
determined by the real eye motion and may be reinforced.

This qualitative description of stochastic mobilevision is sufficient for the
understanding of results as well as their significance for applications but we
need in a more formal definition for their deduction. However, a reader, which
is not interested in formal expositions may omit all mathematical constuctions
below and restrict himself to some comments.

Note once more that to define stochastic mobilevision means to specify its
dynamical perspective laws (dynamical equations, which govern an evolution of
image during observation) and we prefer to do it rather formally in purely
mathematical terms. Such specification is rather analogous to one for the
ordinary MV and is based on concepts of 2D quantum field theory. The
interpretations of mathematical results and their significance for
applications will be commented in detail throughout the text, in the
conclusion and in remarks on applications after it.

\definition{Definition 5} Let $H$ be a canonical projective
$G$--hypermultiplet, $A_t(u,\dot u)$ -- an angular field (obeying the
Euler--Arnold equations $\dot A_t=\{\Cal H,A_t\}$, where the hamiltonian $\Cal
H\in S^{\cdot}(\frak g)$ ($\frak g$ is the Lie algebra of a Lie group $G$) is
a solution of the Virasoro master equation) (or its finite--dimensional
lattice approximations of par.II or [Ju5]). Let $J(u)$ --- an additional
$q_R$--affine current (par.II or [Ju3, Ju4])(or its finite--dimensional
lattice approximation from par.II or [Ju5]) commuting with $G$. A stochastic
evolution equation $$d\Phi(t,[\omega])=A_t(u,\dot
u)\Phi(t,[\omega])\,dt+\lambda J(u)\Phi(t,[\omega])\,d\omega,$$ where
$d\omega$ is the stochastic differential of a Brownian motion (i.e.
$\frac{d\omega}{dt}$ is a white noise), will be called {\it the\/} ({\it
quantum--field\/}) {\it Euler--Belavkin--Kolokoltsov formulas}, the parameter
$\lambda$ will be called {\it the accuracy of measurement\/} (cf. [Be, BK,
Ko1]).
\enddefinition

\remark{Remark 1} These formulas are a reduced version of more general ones
$$d\Phi(t,[\omega])=\{A_t(u,\dot u)+\alpha\lambda^2\,:\!J^2(u)\!:\,\}
\Phi(t,[\omega])\,dt+\lambda J(u)\Phi(t,[\omega])\,d\omega,\qquad (\alpha>0)$$
which will be also called {\it the\/} ({\it quantum--field\/}) {\it
Euler--Belavkin--Kolokoltsov formulas}; $\lambda^2:\!J^2(u)\!:$ is a
Belavkin--type quantum--field counterterm (cf. [Be, BK, Ko1, Ko2]), where
$:\!J^2(u)\!:$ is a spin--2 primary field received from the current $J(u)$ by
the truncated Sugawara construction [Ju3].
\endremark

Here $u=u(t)$ and $\dot u=\dot u(t)$ are the slow variables [Ju2] of
observation (sight fixing point and its relative velocity), the tremor is
simulated by a stochastic differential $d\omega$, $\lambda$ is a {\it free\/}
parameter, $\Phi=\Phi(t,[\omega])\in H$ is a collective notation for a set of
all continuously distributed characteristics of image [Ju2, Ju4, Ju5], $q_R$
is a free internal parameter of a model, which measures the degree of
localization of interaction (the local case corresponds to $q_R=0$). The most
important case is one of $q_R\ll 1$ and all our results will hold for this
region of values of $q_R$. The stochastic Euler--Belavkin--Kolokoltsov
formulas coupled with the deterministic Euler--Arnold equations define a
dynamics, which may be considered as {\it a candidate\/} for one of {\it a
continuously observed\/} ({\it and interactively controlled\/}) {\it
quantum--field top} [Ju3].

\remark{Remark 2} It should be specially emphasized that in stochastic
mobilevision $\lambda$ is a {\it free\/} parameter, which may be chosen
arbitrary by hands (f.e. as great as it is necessary). It means that slow
movements (including saccads) and tremor are decoupled, the firsts are
considered such as in an ordinary MV, whereas the least is stochastized in a
way that {\it its amplitude may be reinforced}.
\endremark

\remark{Remark 3} As it was mentioned above the internal parameter $q_R$
measures a degree of localization of a man--machine interaction in MV and SMV.
It is natural to suppose that the Belavkin--Kolokoltsov watch--dog effects
will appear for sufficiently small values of $q_R$ and the condition $q_R\to 0$
will produce the diminishing of stochastic cores of image. Indeed, we shall see
that sizes of stochastic cores diminish if $q_R$ tends to $0$ and $\lambda$
increases.
\endremark

Below we shall work presumably with finite--dimensional lattice approximations
(cf. [Ko2]) and the associate evolution equation in $H^*$ (see par.II or
[Ju5]), keeping all notations. Also $\Phi$ will be considered as defined on a
compact (the screen of a display or a cluster). It should be marked that in
this case the Euler--Belavkin--Kolokoltsov formulas are transformed into the
ordinary (matrix) stochastic differential equations of diffusion type [GS,
Sk], and hence, $\Phi=\Phi_t=\Phi(t,[\omega])$ is a diffusion Markov process
[Dy].

\remark{Remark 4} Lattice approximations of the ordinary (unobserved and
non--controlled) quantum--field top (in this case angular fields are reduced
to single currents) were actively investigated by St.Petersburg Group directed
by Acad.L.D.Faddeev [AFS]. The main difficulties (technical as well as
principal) in their treatments were caused by a locality of ordinary ($q_R=0$)
affine currents. However, $q_R$--affine currents are not local so their
discretizing is easily performed (see par.II or [Ju5]). It is very interesting
to
receive lattice current algebras of [AFS] from naturally discretized
$q_R$--affine currents by a limit transition $q_R\to 0$, but this problem is a
bit out of the line here.
\endremark

The fact that the ordinary quantum--filed top may be received as a particular
case of our construction ($\lambda=0$, $q_R=0$, $A(u,\dot u)=J(u)\dot u$,
where $J(u)$ is a current) motivates to consider our object as a continuously
observed (and interactively controlled) quantum--field top. Continuous
observation means the inclusion of a stochastic term ($\lambda\ne 0$), whereas
the interactive controlling means the presence of complete algular fields
$A(u,\dot u)=\sum_k B_k(u)\dot u^k$, where $B_k(u)$ are primary fields of spin
$k$, instead of single currents. It seems that these arguments are sufficient
for our terminological innovation.

\remark{Remark 5} The Euler--Belavkin--Kolokoltsov formulas are {\it
postulated\/} to be the dynamical perspective laws for stochastic
mobilevision.  So they are regarded as {\it a mathematical definition of SMV}.
{}From such point of view a transition from MV to SMV consists in:
\roster
\item"1)" the decoupling of slow movements (including saccads) and tremor;
\item"2)" a stochastization of tremor;
\item"3)" the setting the controlling
parameter $\lambda$ free, so that its value may be chosen by hands and it is
not completely determined by real parameters of the eye motions.
\endroster
Thus, the main difference between MV and SMV is that tremor in MV is {\it an
external process\/} governing an evolution of a computer graphic picture,
whereas its stochastization is {\it an internal process\/} (in spirit of {\it
endophysics\/} of Prof.~O.E.~R\"oss\-ler [R\"o, En]) and its characteristics
may be specified by hands. \endremark

Let's summarize the material of par.3.1. Note once more that the ordinary
mobilevision is an interactive computer-graphic system, the evolution of
images in which is governed by the eye movements in accordance to the certain
dynamical perspective laws, which were written in par.II or [Ju2, Ju3, Ju4,
Ju5]. Stochastic mobilevision is an analogous interactive computer-graphic
system, but with slightly different dynamical perspective laws. Namely, in the
dynamical perspective laws of MV the high--frequency eye tremor is decoupled
from the slow eye motions (including saccads), is stochastized in such a way
that it may be considered as {\it purely internal process\/} in the system so
that its characteristics are not completely determined by eye motions and may
be reinforced. So the parameters of an external real process (eye tremor) may
be {\it transformed and scaled up\/} to receive ones of an internal virtual
process (stochastization of tremor). For the understanding of results the
explicit form of dynamical perspective laws is not necessary though it is, of
course, unavoidable for their deduction, which is presented in par.3.2., which
may be omitted by a reader interested only in applications, who may restrict
himself by the comment and remark at its end.

\subhead 3.2. Mathematical analysis\endsubhead
Let $D_A(\Phi)=\left<A^2-\left<A\right>^2_\Phi\right>_\Phi$,
$\left<A\right>_\Phi=\frac{(A\Phi,\Phi)}{(\Phi,\Phi)}$ (Kolokoltsov 1993). It
should be mentioned that one may consider the Euler--Belavkin--Kolokoltsov
formulas with a redefined quantum field $\tilde J(u)=J(u)-\left<J(u)\right>$
instead of the $q_R$--affine current $J(u)$ to receive a full likeness to the
original Belavkin quantum filtering equation [Be, BK, Ko1, Ko2] if the inner
(scalar) product $(\cdot,\cdot)$ is claimed to be translation invariant and
scaling homogeneous. $E_\Phi$ is the mathematical mean with respect to the
standard Wiener measure for observation process with initial point $\Phi$
[Ko2].

\proclaim{Lemma 1}
$$(\forall\Phi_0)\quad\limsup_{t\to\infty}
E_{\Phi_0} D_J(\Phi(t,[\omega]))=
K\lambda^{-2}\longrightarrow_{\lambda\to\infty}0.$$
\endproclaim

The l.h.s. expression (multipled by $\lambda^2$, i.e. just the constant $K$)
is called {\it the Kolokoltsov coefficient\/} of quality of measurement
[Ko2].

\demo{Sketch of the proof} Indeed
$$\aligned\lambda^2\limsup_{t\to\infty}E_{\phi_0}D_J(\Phi(t,[\omega]))=&
\limsup_{t\to\infty}E_{\Phi_0}D_{\lambda J}(\Phi(t,[\omega]))=\\
&\limsup_{t\to\infty}E_{\widetilde\Phi_0}D_J(\widetilde\Phi(t,[\omega])),
\endaligned$$ where $\widetilde\Phi$ is a solution of the
Euler--Belavkin--Kolokoltsov formulas with $\lambda=1$ and with the initial
data $\widetilde\Phi_0$ being equal to $\Phi_0$ scaled in $\lambda$ times (the
least equality follows from the scaling homogenity of the
Euler--Belavkin--Kolkol'tsov formulas). As a sequence of results of [Ko2](the
dependence of the $q_R$--affine current $J$ on $u$ is not essential in view of
the translation invariance) the expression
$\limsup_{t\to\infty}E_{\widetilde\Phi_0}D_J(\widetilde\Phi(t,[\omega]))$,
being the Kolokoltsov coefficient $\kappa(A_t,J)$ for the pair $(A_t,J)$, does
not depend on $\widetilde\Phi_0$, and hence, it is certainly independent on
$\lambda$.
\enddemo

\remark{Remark 6} The sketch of the proof is rather instructive itself.
Instead of difficult calculations of the stationary probability measure (cf.
[Ko2]) and a complicated estimation of its
$\lambda$--behaviour (that is non--trivial to perform rather in the simplest
2--dimensional case considered in [Ko2]) we use general group--theoretical
properties (the translation invariance and the scaling homogenity) of the
Euler--Belavkin--Kolokoltsov formulas, combining them with the strong results
of [Ko2] on an existence of the Kolokoltsov coefficient $K=\kappa(A_t,J)$ and
its independence on the initial data.
\endremark

\remark{Comments on the proof} Concerning the sketch of the proof two remarks
on some details should be made. First, in view of the dependence of the
angular field $A_t(u,\dot u)$ on the controlling parameters the unique
stationary probability measure does not exist; however, we consider all
controlling parameters as slow ones so one may {\sl assume} that there exists
the slowly evoluting stationary probability measure, which form depends only
on the current values of controlling parameters (of course, it is clear that
such assumption is natural from mathematical physics point of view, however,
it means a certain {\it "gap"\/} in the rigorous proof from pure mathematics
one; but here any {\it "purification"\/} will be out of place). Such
parameters varies through a compact set (in the continuous version, or may
have only finite number of values in the lattice version), so one can define
the Kolokoltsov coefficient as the supremum of such coefficients calculated
for the measures from the compact (or finite) set (just this circumstance
causes the appearing of "$\limsup$" in Lemma 1). However, second, now one may
use the scaling rigorously only for infinite regions, whereas we have to deal
with finite ones (the screen of a display or clusters); however, the
transition to the compact regions may only cause that the Kolokoltsov
coefficient $K$ being a function of $\lambda$ decreases if $\lambda$ tends to
infinity.
\endremark

Let's $Q$ be the coordinate operator $Qf(x)=xf(x)$; $J^\circ$ be a singular
part of the current $J$ (par.II or [Ju2, Ju5]), i.e. $J^\circ(u)=(Q-u)^{-1}$.

\proclaim{Lemma 2}
$$E_{\Phi_0}
\left(D_J(\Phi(t,[\omega]))-D_{J^\circ}(\Phi(t,[\omega]))\right)
\rightrightarrows 0\quad\text{ if }\quad q_R\to 0. $$
\endproclaim

It should be marked that the statement of the lemma na\"\i vely holds only in
the continuous version; after a finite--dimensional approximation the
expression "$\rightrightarrows 0$" should be understand as the l.h.s. becomes
uniformely less than a sufficiently small constant $\epsilon$ (which depends
on the chosen approximation), when $q_R$ tends to zero.

\demo{Hint to the proof} The lemma follows from the explicit computations of
eigenfunctions of a $q_R$--conformal current $J(u)$.
\enddemo

\proclaim{Main Theorem}
$$(\forall\Phi_0)\quad\lim_{\lambda\to\infty, q_R\to 0}\limsup_{t\to\infty}
E_{\Phi_0} D_Q(\Phi(t,[\omega]))=0.$$
\endproclaim

The statement of the theorem is a natural sequence of two lemmas above; it
remains true in the multi--user mode [Ju6] also. Certainly, the
statement of the theorem na\"\i vely holds only in the continuous version (cf.
Lemma 2); after a finite--dimensional approximation the equality of the limit
to 0 should mean that this limit is less than a sufficiently small constant
$\epsilon$, which depends on the chosen approximation.

\remark{Comment} Thus, we received that the Belavkin--Kolokoltsov watch--dog
effects in stochastic mobilevision appear for all values of the accuracy of
measurement $\lambda$ for sufficiently small values of parameter $q_R$.
Moreover, if $\lambda$ increases and $q_R$ tends to $0$ the stochastic cores
may be diminished to several pixels.
\endremark

\remark{Remark 7} Note that the Belavkin--Kolokoltsov watch--dog effects
appear only in the models of SMV with sufficiently small values of the
internal parameter $q_R$, which measures the localization of interaction
($q_R=0$ mens the local case). However, $q_R$, being an internal parameter,
may be chosen in arbitrary way, so the condition $q_R\ll 1$ may be always
provided.
\endremark

\head IV. CONCLUSION\endhead

\subhead 4.1. Summary of results\endsubhead
Thus, the results may be briefly summarized.

First, let's emphasize once more that the main difference of SMV from the
ordinary MV is that the stochastization of eye tremor in the first is
considered as an internal process, so its amplitude characteristics may be
{\it reinforced}. Second, for {\it all values\/} of $\lambda$ (a free
parameter of such stochastization, which measures the reinforcing of the
amplitude of tremor --- the so--called accuracy of measurement) the
Belavkin--Kolkoltsov watch--dog effects for stochastic dynamics of image in
SMV are observed (it means that stochastic cores of image have finite sizes
for all times) for sufficiently small values of an additional internal
parameter $q_R$; it confirms the presence of watch--dog effects also in the
models of ordinary MV with the same $q_R$. Moreover, third, if the value of
$\lambda$ is great enough, whereas $q_R\ll 1$ than the stochastic scores of
SMV image may be diminish to several pixels. Such effect, which is produced by
the reinforcing of $\lambda$, may be effectively used in practical
computer-graphics for various purposes as it was marked in the introduction.
Some further discussions of significance of the obtained results for other
applications may be found in par.4.2.

\subhead 4.2. Remarks on applications\endsubhead
Besides theoretical importance for the
interactive visualization of 2D quantum field theory the results of the paper
seems to be useful for applications to ({\bf 1}) the elaboration of
computer-graphic interactive systems for psychophysiological self--regulation
and cognitive stimulation [Ju4, Ju5], ({\bf 2}) the interactive
computer-graphic modelling of a "quantum computer" [Ju5] (see [D, Jo,
DJ] for a general discussion on "quantum computers" and their use for rapid
computations as well as [Un] on fundamental difficulties to construct the
"physical" non-interactive "quantum computer"), which may be used for an
actual problem of the accelerated processing of the complex sensorial data in
the "virtual reality" (visual--sensorial) networks, ({\bf 3}) the creation of
computer graphic networks of tele\ae sthetic communication [Ju5].

Let's discuss a significance of obtained results for these applications.

\remark{Comment: Obtained results and applications}

({\bf 2}) is directly related to our results because the maintaining of the
coherence is the main problem for "quantum computers". As it was mentioned
earlier [Ju5] MV may be regarded as an interactive computer-graphic
simulation of a "quantum computer" behavior. The presence of free parameters
(such as $\lambda$) in SMV allows to maintain the coherence for long times with
an arbitrary precision in the interactive mode.

Moreover, such interactive computer-graphic simulations may be more useful
than the original "quantum computers" for the "virtual reality" problems in
view of the implicit presence of graphical datain the interactive mode. A
reorganization of these data by the secondary image synthesis [Ju7]
and their representation via MV or SMV may allow an accelerated parallel
processing of the complex sensorial data in such systems.

({\bf 1}) and ({\bf 3}) are indirectly related to our results because they
depend on a solution of the main problem of dynamics in interactive
psychoinformation computer-graphic systems (a problem of the nondemolition of
images). For ({\bf 3}) its solution allows to transmit the graphically
organized information without a dissipation and additional errors. For ({\bf
1}) its solution allows to consider a long--time self--organizing interactive
processes, which play a crucial role in systems for psychophysiological
self--regulation and cognitive stimulation.
\endremark

So it should be stressed that the obtained results are essential for the
prescribed applications.

\subhead 4.3. Remarks on generalizations and perspectives\endsubhead
Now let's discuss the possible generalizations.

Really one consider a random (discrete) simulation of the continuous Brownian
motion and stochastic differentials.  It may be rather interesting to replace
it by any their perturbation (f.e. by some version of the weakly
self--avoiding or self--attracting walks, especially by their finite memory
approximations).

First, these generalizations are motivated by the fact that Brownian motion
may be not the best stochastization of the eye tremor. Really, it may be
considered only as a first approximation for tremor, whereas the more
complicated models will be preferable. However, it seems that the watch--dog
effects are conserved by any form of the weakly self--attracting
perturbations, which are the most realistic candidates for tremor.

Second, it seems to be rather interesting to use the decoupling of
high--frequency tremor from slow eye movements (including saccads) and an
internal character of its stochastic simulations for the organization of
various "intelligent" forms of human--computer interaction (the so--called
{\it "semi--artificial intelligence"\/}). In such approach the stochastized
tremor plays a role of an internal observer (cf. [R\"o, En]), which presence
is crucial for a self--organization of graphical data in systems of the
semi--artificial intelligence [KT]. But this topic (though being related to
({\bf 1}) above) seems to be too manysided and too intriguing that this paper
is not a suitable place to discuss it further.

\head ACKNOWLEDGEMENTS
\endhead

The author is undebtful to Prof.Dr.V.N.Kolokoltsov (Applied Mathematics
Department, Moscow Institute for Electronics and Mathematics (MIEM), Moscow,
Russia) for an attention and numerous discussions as well as to Laboratoire de
Physique Th\'eorique de l'\'Ecole Normale Sup\'erieure for a kind hospitality
and a beautiful atmosphere.

The author also thanks his referees for very valuable comments, constructive
remarks and useful suggestions.


\Refs
\widestnumber\key{ZMNP}
\ref\key AFS\by Alekseev, A., Faddeev, L., Semenov--Tian--Shansky, M.\paper
Hidden quantum groups inside Kac--Moody algebra\jour Commun.~Math.~Phys.\yr
1992\vol 149\pages 335--345\endref
\ref\key BC\by Burdea, G., Coiffet, Ph.\book Virtual reality
technology\publaddr
New York\publ John Wiley \&\ Sons\yr 1994\endref
\ref\key Be\by Belavkin, V.P.\paper Nondemolition measurement, nonlinear
filtering and dynamic programming of quantum stochastic processes\jour
Lect.~Notes Contr.~Inform.~Sci.\yr 1988\vol 121\endref
\ref\key BJ1\by Bychkov, S.A., Juriev, D.V.\paper Fubini--Veneziano fields in
quantum projective field theory\jour Uspekhi Matem.~Nauk\vol 46\issue 5\yr
1991\pages 167--168\lang in Russian\endref
\ref\key BJ2\by Bychkov, S.A., Juriev, D.V.\paper Three algebraic structures of
quantum projective field theory\jour Theor.~Math.~Phys.\yr 1993\vol 97
\pages 1333--1339\endref
\ref\key BK\by Belavkin, V.P., Kolokoltsov, V.N.\paper Quasiclassical
asymptotics
of quantum stochastic equations\jour Theor.~Math.~Phys.\yr 1991\vol 89\pages
1127--1138\endref
\ref\key By\by Bychkov, S.A.\paper Fubini--Veneziano fields in projective Verma
quasibundles\jour Uspekhi Ma\-tem. Nauk\vol 47\issue 4\yr 1992\pages
187--188\lang in Russian\endref
\ref\key CSM\by Chiu, C.B., Sudarshan, E.C.G., Misra, B.\paper Time evolution
of
unstable quantum states and a resolution of Zeno's paradox\jour Phys.~Rev.~D
\yr 1977\vol 16\pages 520--529\endref
\ref\key Do\by Deutsch, D.\paper Quantum theory, the Church--Turing principle
and
the universal quantum computer\jour Proc.~Royal Soc.~A\yr 1985\vol 400\pages
97--117\endref
\ref\key DJ\by Deutsch, D., Jozsa, R.\paper Rapid solution of problems by
quantum
computation\jour Proc.~Royal Soc.~A\yr 1992\vol 439\pages 553--558\endref
\ref\key Dy\by Dynkin, E.B.\book Markov processes\publ Springer--Verlag\yr
1965\endref
\ref\key En\by {\it Endophysics\/}\book\nofrills\eds P.Weibel, O.E.R\"ossler,
G.Kampis\publ Aerial Press\yr 1993\endref
\ref\key GS\by Gihman, I.I., Skorohod, A.V.\book The theory of stochastic
processes. III\publ Springer--Verlag\yr 1979\endref
\ref\key Jo\by Jozsa, R.\paper Characterizing classes of functions computable
by
quantum parallelism\jour Proc. Royal Soc.~A\yr 1991\vol 435\pages
563--574\endref
\ref\key Ju1\by Juriev, D.\paper Classification of vertex operators in
two--dimensional quantum $\slt$--in\-va\-ri\-ant field theory\jour
Theor.~Math.~Phys.\yr 1991\vol 86\pages\endref
\ref\key Ju2\by Juriev, D.\paper Quantum projective field theory:
quantum--field
analogs of Euler formulas\jour Theor.~Math.~Phys.\yr 1992\vol 92\pages
814--816\endref
\ref\key Ju3\by Juriev, D.\paper Quantum projective field theory:
quantum--field
analogs of Euler--Arnold equ\-a\-tions in projective $G$--hypermultiplets\jour
Theor.~Math.~Phys.\yr 1994\vol 98\pages 147--161\endref
\ref\key Ju4\by Juriev, D.\paper Octonions and binocular mobilevision\jour
E-print (LANL archive on Theor.~High Energy Phys.): {\it hep-th/9401047}\yr
1994\endref
\ref\key Ju5\by Juriev, D.\paper Visualizing 2D quantum field theory: geometry
and informatics of mobilevision\jour E-print (LANL archive on Theor.~High
Energy Phys.): {\it hep-th/9401067}\yr 1994\endref
\ref\key Ju6\by Juriev, D.\paper The advantage of a multi--user mode\jour
E-print (LANL archive on Theor.~High Energy Phys.): {\it hep-th/9404137}\yr
1994\endref
\ref\key Ju7\by Juriev, D.\paper Secondary image synthesis in electronic
computer
photography\jour E-print (LANL archive on Adap.~Self-Org.~Stoch.): {\it
adap-org/9409002\/}\yr 1994\lang in Russian\endref
\ref\key Ju8\by Juriev, D.\paper Belavkin--Kolokoltsov watch--dog effects in
interactively controlled stochastic computer-graphic dynamic systems. A
mathematical study\jour E-print (LANL archive on Chaos \&\ Dyn.~Systems):
{\it chao-dyn/9406013}\yr 1994\endref
\ref\key KT\by Kaneko, K., Tsuda, I.\paper Constructive complexity and
artificial
reality: an introduction\jour E-print (LANL archive on Adap.~Self-Org.~Stoch.):
{\it adap-org/9407001}\yr 1994\endref
\ref\key Ko1\by Kolokoltsov, V.N.\paper Application of the quasiclassical
methods
to the investigation of the Belavkin quantum filtering equation\jour
Math.~Notes\yr 1991\vol 50\pages 1204--1206\endref
\ref\key Ko2\by Kolokoltsov, V.N.\paper Long time behavior of continuously
observed and controlled quantum systems (a study of the Belavkin quantum
filtering equation)\jour Preprint Institut f\"ur Mathematik,
Ruhr-Universit\"at-Bochum\yr 1993\issue 204\endref
\ref\key MF\by Mi\v s\v cenko, A.S., Fomenko, A.T.\paper Euler equations on
finite dimensional Lie groups\jour Is\-ves\-tiya AN SSSR, Ser. Math.\yr
1978\vol 12\pages 371--390\lang in Russian\endref
\ref\key MS\by Misra, B., Sudarshan, E.C.G.\paper The Zeno's paradox in quantum
theory\jour J.~Math.~Phys.\yr 1977\vol 18\pages 756--763\endref
\ref\key Rh\by Rheingold, H.\book Virtual reality\publ Summit Books\publaddr
New York, Tokyo\yr 1991\endref
\ref\key Ro\by Ro\v cek, M.\paper Representation theory of the nonlinear
$\SU(2)$ algebra\jour Phys.~Lett.~B\yr 1991\vol 255\pages 554-557\endref
\ref\key R\"o\by R\"ossler, O.E.\paper Endophysics\inbook Real brains,
artificial
minds\eds J.L.Carti \&\ A.Karlqvist\publ  North Holland\yr 1987\endref
\ref\key Sk\by Skorohod, A.V.\paper Operator stochastic differential equations
and stochastic semigroups\jour Russian Math.~Surveys\yr 1982\vol 37\issue 6
\pages 177--204\endref
\ref\key SV1\by {\it Scientific visualization of physical phenomena\/}\book
\nofrills\ed N.M.Patrikalakis\publ Springer--Verlag\yr 1991\endref
\ref\key SV2\by {\it Scientific visualization: techniques and applications\/}
\book\nofrills\ed K.W.Brodlie\publ Springer--Ver\-lag\yr 1992\endref
\ref\key Un\by Unruh, W.G.\paper Maintaining coherence in quantum
computers\jour
E-print (LANL archive on Theor.~High Energy Phys.): {\it hep-th/9406058}\yr
1994\endref
\ref\key Vi1\by {\it Visualization in human--computer interaction\/}\book
\nofrills\eds P.Gorny, M.J.Tauber\publ Springer\yr 1990\endref
\ref\key Vi2\by {\it Visualization '93\/}\book\nofrills\publaddr Los Alamitos,
Calif.\publ IEEE Computer Society Press\yr 1993\endref
\ref\key VR1\by{\it Virtual realities\/}\inbook Visual computing\ed T.L.Kunii
\publ Springer--Verlag\yr 1992\endref
\ref\key VR2\by {\it Virtual reality: applications and explorations\/}\book
\nofrills\ed A.Wexelblat\publaddr Boston\publ Acad. Publ.\yr 1993\endref
\ref\key VR3\by {\it Virtual reality: an international directory of research
projects\/}\book\nofrills\ed J.Thompson\publaddr Westport\publ Meckler\yr
1993\endref
\ref\key ZMNP\by Zakharov, V.E., Manakov, S.V., Novikov, S.P., Pitaevskii,
L.P.\paper Soliton theory: inverse scattering problem method\publaddr
New York\publ Consultants Bureau\yr 1984\endref\endRefs
\enddocument